\documentstyle[prl,aps,epsfig,preprint]{revtex}

\draft

\begin{document}                % INITIALIZE - DONT CHANGe

\title{Coherent Control of Atom-Atom Interactions and Entanglement using Optical 
Fields}
\author{M.~D.~Lukin$^1$ and P.~R.~Hemmer$^2$}
\address{$^1$ ITAMP, Harvard-Smithsonian Center for Astrophysics, Cambridge,
MA 02138, \\
$^2$ Sensors Directorate, Air Force Research Laboratory, Hanscom Air Force Base, Bedford, MA 01731}

\date{\today}

\maketitle

\begin{abstract}

Two-photon optical transitions combined with 
long-range dipole-dipole interactions can be used for the
coherent manipulation of collective metastable states composed of
different atoms. We show 
that it is possible to induce optical resonances 
accompanied by the generation of 
entangled superpositions of the atomic states. Resonances of this 
kind can be used to implement quantum logic gates using optically excited 
single atoms (impurities) in the condensed phase.   
\end{abstract}


\newpage

Exciting recent developments in the field of quantum information and quantum
computing stimulated an intensive search for coherent physical processes which 
could be used to manipulate coupled quantum-mechanical systems in 
a prescribed fashion \cite{qcomp}. Although it is clear that  so-called 
universal  
quantum-mechanical computers are well beyond the abilities of current 
technology, even small-scale devices consisting of few interacting quantum 
bits are currently of significant fundamental interest.  
 
The present Letter describes a new method for coherent manipulation 
of many-atom metastable states based on two-photon optical 
resonances in interacting atomic ensembles. 
Such collective resonances take place due to the coupling of the 
optical dipoles on selected, field-free transitions 
via a dipole-dipole interaction. 
Even though the mean values of the individual dipole moments involved in 
such atom-atom interactions vanish, this  coupling can 
result in ``hopping'' of the excitation from one atom to another. Splitting 
of collective excited states due to this ``hopping'' is a dominant 
feature of the present problem. In such a 
system it is possible to induce a narrow two-photon transition between 
two distinct atoms or to perform a two-photon transition within one atom 
conditional upon the metastable state of the other. When excited with a 
prescribed pulse sequence these resonances can be used for the generation 
of entangled superpositions of metastable atomic states and conditional 
quantum logic.  As a specific example we 
suggest a solid state implementation of a quantum computer, in which individual
impurities or defects with long lived  ground state coherences can be excited
optically. 

Substantial progress has been made in recent years towards the understanding of
the interactions of optical fields with dense ensembles of 
multi-state atoms \cite{steve2}.   
Particularly relevant studies of solid materials  
\cite{hole} and ultra-cold atoms \cite{lene} should be noted.
Dipole-dipole induced ``hopping'' or transfer of the excitation between 
different atoms has been 
extensively studied in the past \cite{fost} and its manifestations have 
been observed e.g. in a dense thermal vapor \cite{sau}.  Recent work 
utilizing these effects for ``nano-optics'' \cite{let} might also have 
interesting implications for the experimental realization of the present ideas.
Before proceeding we note that adiabatic following  in ``dark states'' 
is the basis for a cavity QED-based  quantum logic as described in Ref. \cite{zoler}. We also note very recent
related proposals for quantum logic, based on the precise control of atomic positions in optical 
lattices \cite{recent1,recent2}.  
 
Consider a pair of  distinguishable  multi-state atoms (A and B)
at a fixed distance $r_{AB}$ which is smaller than the  optical 
wavelength $\lambda$ 
(Fig.1a). Each of these atoms is assumed to have a single excited state, which 
can be coupled to several metastable states via electric dipole-allowed 
transitions. 
We assume that the atoms can only interact with each other via the coupling of 
dipoles on selected transitions ($|a_i\rangle \rightarrow |b_i\rangle$, 
$i=A,B$ in Fig.1a),
whereas   all the other dipoles (e.g. $|a_i\rangle \rightarrow |c_i\rangle$) 
do not couple to those in other atoms due to  
differing frequencies or polarizations. In the situations considered below, 
the coupled dipoles are assumed not to be driven by any external 
electromagnetic field, and  we will examine the response of this type of 
system when 
optical fields are applied to uncoupled transitions. In cases when 
retardation effects can be disregarded the system can be described by 
the following effective Hamiltonian:
\begin{eqnarray}
H &=& \sum_{i=A,B} (H^a_i + V^{a-f}_i) - (\hbar g(\vec{r}_{AB}) 
\sigma_{ab}^A \sigma_{ba}^B + {\rm h.c.}), 
\end{eqnarray}
where $H^a_i$ corresponds to free atoms, $V^{a-f}_i$ describes 
interaction of each atom with components of electromagnetic field, 
$\sigma_{\alpha\beta}^{i} = |\alpha_i\rangle \langle \beta_i|$ are 
dipole operators and $g(\vec{r}) = \hbar^{-1} \wp^A_{a\rightarrow b} 
\wp^B_{a\rightarrow b}/r^3 (3 z^2/r^2 -1) = 3/2 (2 \pi)^{-3} \sqrt{\gamma^A_{a\rightarrow b}
\gamma^B_{a\rightarrow b}} \lambda^3/r^3 (3 z^2/r^2 -1)$  is the coupling 
constant between interacting dipoles. Here $\wp^k_{i\rightarrow j}$ are single-atom
dipole matrix elements corresponding to  $i\rightarrow j$ transition of the
kth atom, and $\gamma^k_{i\rightarrow j}$ are corresponding radiative decay
rates.

Let us first consider the case when each of the three-state 
atoms $i=A,B$ in Fig.1a is coupled by an independent optical field with a respective 
Rabi-frequency $\Omega_{1,2}$ ($\nu_{1,2}$ are oscillation 
frequencies), as shown. These fields are tuned to 
resonance with the transitions $|a_i\rangle \rightarrow |c_i\rangle$ and 
atoms $A$ and $B$ are initially prepared in their lowest metastable states $c_A$ 
and $b_B$, respectively.  This configuration corresponds to a Raman transition 
between two different atoms which will result in the level $c_A$ being emptied while level $c_B$ is filled. To get an insight into the origin of this transition let us 
consider collective energy levels of the two-atom system  
dressed by coherent driving field $\Omega_2$ (Fig.1b). 
Three relevant dressed states $|0\rangle,|\pm \rangle$ 
with energies 
$\omega_0,\omega_{\pm}$ can be excited by the probe field $\Omega_1$. 
In the simple case of 
an exactly resonant driving field $\Omega_2$, close atoms $|g| \gg \Omega_2$  
and 
interacting transitions $a_i\rightarrow b_i$ of equal energies 
($\omega^A_{ab} =\omega^B_{ab}$) we find
\begin{eqnarray}
\omega_0 &=& 0, \quad |0 \rangle \approx {-\Omega_2 |a_Ab_B\rangle + 
g|b_Ac_B\rangle \over |g|} \nonumber \\
\omega_\pm &\approx& \pm |g|, \; |\pm \rangle 
\approx {|a_Ab_B \rangle \pm |b_Aa_B\rangle
\over \sqrt{2}} + O({|\Omega_2| \over |g|}). \nonumber
\end{eqnarray}  
The dominant feature of the present problem is splitting of the
states $|\pm\rangle$.  
The third dressed state $|0\rangle$ coincides nearly identically with the
metastable two-atom state $|b_Ac_B\rangle$.
At the same time  
it contains a small admixture of the state $|a_Ab_B\rangle$
and therefore can be excited from the initial state 
$|c_A b_B\rangle$ by the 
field $\Omega_1$ resulting in the two-atom Raman transition. 

These effects can be quantified by computing the response of the
two-atom system assuming a cw driving field $\Omega_2$ 
and a weak probe field $\Omega_1$. For a moment we disregard cooperative 
relaxation (assuming that dephasing dominates in the relaxation of the optical 
coherences) and we find for the expectation value of the induced 
atomic polarization:
\begin{eqnarray}
\langle \sigma_{ac}^A \rangle = i \Omega_1 
 {\Gamma_{bcab} \Gamma_{cbbc} + |\Omega_2|^2 \over 
\Gamma_{ac}(\Gamma_{bcab} \Gamma_{bccb} + |\Omega_2|^2) + g^2 \Gamma_{bccb}}.
\label{pol1}
\end{eqnarray}  
Here $\Gamma_{ac} = \gamma^A_{ac} + i (\nu_1 - \omega^A_{ac})$,  
$\Gamma_{bcab} = \gamma^A_{bc} + \gamma^B_{ab} + i (\nu_1 - 
\omega^A_{cb}-\omega^B_{ab})$, and 
$\Gamma_{bccb} = \gamma^A_{bc} + \gamma^B_{cb} + 
i (\nu_1 - \nu_2 - \omega^A_{bc} - \omega^B_{cb})$ are complex 
relaxation rates of a polarization $ 
\langle \sigma_{ac}^{A} \rangle$ and coherences  
$\langle \sigma_{bc}^{A} \sigma_{ab}^{B} 
\rangle$, and  $\langle \sigma_{bc}^{A} \sigma_{cb}^{B}\rangle$. 
$\gamma^k_{ij}$ are the coherence decay rates for the respective 
transitions of the individual atoms.  The susceptibility spectrum 
corresponding to Eq. (\ref{pol1}) is shown in Fig.1c.  
Note that if the $|\pm\rangle$ dressed state splitting caused
by the atom-atom interaction $\sim 2 |g|$ 
is larger that the linewidth of the 
atomic transitions ($\gamma$), the narrow 
$|c_Ab_B\rangle \rightarrow |0\rangle$
resonance (solid curve in Fig.1c) can be selectively excited
without significantly populating the upper states. Excitation 
of this resonance with pulsed fields results  in a 
a metastable, entangled  superposition  
$|\Psi_{final} \rangle = 
\xi_{cb} |c_Ab_B\rangle + \xi_{bc} |b_Ac_B\rangle$. Here  
$\xi_{ij}$ are complex numbers determined by Rabi frequencies, 
pulse lengths and magnitude of $g$ \cite{adiab}.

It is easy to generalize the above analysis to the case when 
the atom $A$ is initially in coherent superposition $|A\rangle = \alpha 
|b_A\rangle + \beta |c_A\rangle$ and $B$ is in the pure state $|b_B\rangle$ . 
The two-atom state  $|b_Ab_B\rangle$  
remains decoupled from the optical fields; hence 
$|\Psi_{final} \rangle = \alpha 
|b_Ab_B\rangle + \beta (\xi_{cb} |c_Ab_B\rangle + \xi_{bc} |b_Ac_B\rangle)$.
The particular case ($\xi_{cb}\rightarrow 0,\xi_{cb}\rightarrow 1$) 
corresponds to a two-atom Raman $\pi$ pulse that yields complete coherence 
transfer. The inset to Fig.1c illustrates Raman-type Rabi oscillations between 
the metastable states $|c_Ab_B\rangle$ and $|b_Ac_B\rangle$  
corresponding to the transfer of coherence $\sigma_{bc}$ between two atoms.

We next extend the above technique to illustrate how it 
can be used to perform a conditional manipulation of ground state coherences.
To this end (Fig.2a) we provide atoms $A$ and $B$ with two or more 
ground state sublevels that are not coupled to other atoms.  On atom $A$ 
only, we also provide a pair of optical fields $\Omega_{1,2}$ which are 
near resonance with the uncoupled transitions
$|c_A \rangle \rightarrow |a_A \rangle$ and $|d_A \rangle \rightarrow 
|a_A \rangle$ respectively, forming a resonant Raman system, as shown in 
Fig.2a. As 
before, atom $A$ can interact with atom $B$ via the dipole-dipole coupling 
on the $|a_i\rangle \rightarrow |b_i\rangle$ 
transitions.  We assume that the multi-state atom $B$ is initially prepared 
in a coherent superposition of the metastable 
states (e.g. $|B \rangle = \alpha_B |b_B\rangle + \beta_B |c_B\rangle$) and 
the fields do not drive any transitions in this atom.  If 
atom $B$ is in the state $|c_B\rangle$, then its presence does not 
affect the two photon transition within atom $A$ (see Fig.2b).  
However, if atom $B$ is in the state  $|b_B\rangle$ the dipole-dipole coupling 
causes an effective splitting of the excited state (into 
$|\pm \rangle$ states)
resulting in a substantial slowing of 
the two-photon processes (Fig.2c). In such a case, Raman transitions in atom 
$A$ can be effectively eliminated by choosing appropriate Rabi-frequencies 
and pulse durations. 

Two particular examples of such interactions are  
conditional adiabatic passage and conditional Raman excitation with 
resonant pulses.  In the former case, one of the ground states 
(say $|d_A\rangle$) of atom A must be initially unoccupied.      
{\it Conditional adiabatic passage} can be achieved by
applying resonant pulses in the 
form $\Omega_1 = \Omega {\rm cos}(t/T), \Omega_2 = \Omega {\rm sin}(t/T)$ with
properly chosen $\Omega$ and $T$. In the ideal limit, the state of the system 
after the transfer $t = T \pi/2$ is governed by the operator 
$U_{cAP}$: 
\begin{eqnarray}
U_{cAP}(\alpha_B |c_A b_B\rangle + \beta_B |c_A c_B\rangle) = 
\alpha_B |c_A b_B\rangle + \beta_B 
|d_A c_B\rangle. \nonumber
\end{eqnarray}
A {\it conditional Raman $\pi$-pulse} can be achieved by applying 
identically  shaped 
resonant pulses $\Omega_1 (t) = {\bar \Omega}_1 f(t)$, $\Omega_2 (t) = 
{\bar \Omega}_2 f(t)$ with properly chosen intensities and 
durations $\int f(t) dt {\bar \Omega} = \pi$.  
For an arbitrary initial state the ideal limit corresponds to
the unitary operation $U_{c\pi}$:
\begin{eqnarray}
U_{c\pi} |c(d)_A c_B \rangle &=& \pm {\rm cos}  \theta  |c(d)_A c_B \rangle + 
e^{\mp i \phi} {\rm sin} \theta |d(c)_A c_B \rangle, \nonumber\\
U_{c\pi} |c(d)_A b_B \rangle &=& |c(d)_A b_B \rangle, \nonumber
\end{eqnarray}
where ${\rm tan} \theta = 2 |\Omega_1||\Omega_2|/(|\Omega_1|^2 - |\Omega_2|^2)$, $ {\bar \Omega} = \sqrt{|{\bar \Omega}_1|^2 + |{\bar \Omega}_2|^2} $
and $e^{i\phi} = \Omega_1^*\Omega_2/(|\Omega_1||\Omega_2|)$.

As a figure of 
merit for such  processes we compute minimal fidelity $F = 
{\rm min} \; \langle\Psi_f|
\rho_f |\Psi_f\rangle$, where $\rho_f$ is the actual atomic density
 matrix after the pulse sequence and $|\Psi_f\rangle$ is a target 
state achieved in the
ideal limit. Disregarding dephasing of the lower level coherence during 
adiabatic transfer and assuming $\omega_{ab}^A = \omega_{ab}^B$ 
we find that for $|g| > \gamma$ the fidelity of adiabatic passage:
\begin{eqnarray}
F_{cAP} \approx {\rm min} \; [{\rm exp}(-{\pi \gamma \over 2 \Omega^2 T}),
{\rm exp}(-{\pi \gamma \Omega^2 T \over 4 |g|^2})]
\label{fidelity}
\end{eqnarray}
In this expression, the first term corresponds to imperfect adiabatic 
passage and the second corresponds to unwanted transfer. 
It follows that the fidelity is large when $\Omega^2 T \gg \gamma$ 
and $\Omega^2 T \ll |g|^2/\gamma$ (see Fig.3a). A corresponding 
fidelity for 
Raman excitation induced by a pair of rectangular pulses with identical 
Rabi-frequencies is: 
\begin{eqnarray}
F_{c\pi} \approx {\rm min} \;[{\rm exp}(-{\pi \gamma \over 2 {\bar \Omega}}),
{\rm exp}(-{\pi \gamma {\bar \Omega} \over  |g|^2}) ]. 
\label{fidelitypi}
\end{eqnarray}
$F_{c\pi}$
is large when ${\bar \Omega} \gg \gamma$ and ${\bar \Omega} \ll |g|^2/\gamma$.
It follows from Eqs.(\ref{fidelity},\ref{fidelitypi}) that 
whenever the necessary inequalities on pulse durations and Rabi-frequencies 
are fulfilled {\it precise knowledge of magnitude of the dipole 
coupling is not required} to achieve desired conditional manipulations. 
This is especially remarkable since the above 
operations can be used to implement  universal logic gates 
such as a controlled-NOT \cite{cstir}.

The above analysis disregards the effects of cooperative 
relaxation such as superradiance. This is justified since the 
effect of cooperative relaxation 
on collective two-photon transitions is not significant even if the optical 
dephasing is dominated by radiative decay (see e.g. the inset to 
Fig.3b which shows the two-atom spectrum when cooperative relaxation 
is taken into 
account).  In general, cooperative relaxation 
effects are typically dominated by 
the dipole-dipole interaction in dense small-size 
atomic samples \cite{haroche}.

It is possible to  generalized the above analysis
to the case of ensembles of interacting 
atoms consisting of two groups of atoms. 
Fig.3b shows an example of an absorption spectrum corresponding to a Raman 
transition between two atomic ensembles in 
the case of a homogeneous medium consisting of 
 atoms  $A$ and $B$ with equal densities \cite{bb}.  It is clear that 
simultaneous many-particle 
interactions introduce shifts and asymmetry (resulting from effects
such as local field correction \cite{chuck}) as well as additional broad 
absorption plateaus in between the split peaks. The origin  of the latter
can be understood as an emerging exciton band. It is important to note
however that a narrow resonance corresponding to a many-atom Raman transition 
can still be easily resolved, although its magnitude might be reduced.

In summary, collective multiphoton resonances 
should be observable in systems where the shifts due to the
interaction between atoms exceeds the homogeneous optical linewidths and 
where there are several metastable states available. Systems 
of that kind include certain solid materials such as spectral hole 
burning materials and cold atoms in small traps and optical lattices 
\cite{lat}. 

As an important example  we now 
describe a possible implementation of a quantum logic gates using the 
above techniques on single atoms (i.e. impurities or defects) in solid 
crystals. We here take optical spectral hole burning solids as an example.  
Possible candidate materials  include 
Pr doped Y$_2$SiO$_5$ (Pr:YSO) \cite{YSO}, for which efficient Raman 
excitation has been realized \cite{hole}, 
color centers such as the N-V center in diamond (N-V) \cite{NV} for 
which single-atom spectroscopy have been performed \cite{singl}, F$_2$ centers 
in LiF \cite{let} or Cs in solid helium \cite{cesium}.
The ground state of such  materials consists of multiplets of  
degenerate sublevels.  When cooled to liquid  helium temperatures, 
such impurities can display 
homogeneous optical linewidths which are close to radiative broadening 
($\gamma \sim 3$ kHz (Pr:YSO) - $20$ MHz (N-V), 
$\gamma_{rad} \sim \gamma/(2 ... 5) $).  
and relatively long lived ground state coherence lifetimes 
($(\pi \gamma_0)^{-1}\sim 1$ ms (Pr:YSO)  - $0.1$ ms (N-V) \cite{t1}).

Due to crystal field fluctuations there is a large ($\Delta_{inh}\sim $ 
5 GHz (Pr:YSO) - few THz (N-V)) inhomogeneous distribution of the optical 
frequencies.  
This allows one to use spectral hole burning techniques  
to select atoms with optical transitions at desired frequencies. 
In these systems qbits can be defined as pairs of ground state 
atomic sublevels corresponding to spectrally selected atoms \cite{phil}. 
All but relevant atoms can be optically pumped into designated storage 
sublevels. In the current approach we choose to 
excite samples  of a very small ($< \lambda$) size by using near-field
microscopy or tightly focused beams, and choose the density such 
that each spectral hole of interest 
consists  of only one impurity atom. In this case
it is possible to selectively address single atoms  
\cite{singl}. 
A pair of individual impurity atoms $A$ and $B$
can be selectively coupled by shifting  the Zeeman 
sublevels with an H-field until one optical transition 
  of the atom $A$  becomes resonant with another transition of 
the atom $B$. Zeeman frequency shifts in the above mentioned impurities
are such that magnetic fields on the order of few Tesla are appropriate. 
Once such ``alignment of levels'' is achieved, the two-atom multiphoton 
transitions described above can be used to perform quantum logic operations 
on a time scale of at least  $\gamma^{-1}$.
By choosing an appropriate concentration of impurities and sample size 
the excitation and a few steps of coherent manipulation 
of about hundred of qbits might be feasible.

The authors gratefully acknowledge many 
useful discussions with M. Fleischhauer,  V. Kharchenko and 
V. Sautenkov. This work was supported by the 
National Science Foundation.

%\newpage

%%%%%%%%%%%%%%%%%%%%%% figure 1 %%%%%%%%%%%%%%%%%%%%%%%%%%%%%%%%%%%%%%%%%%%%%

\begin{figure}[ht]
 \centerline{\epsfig{file=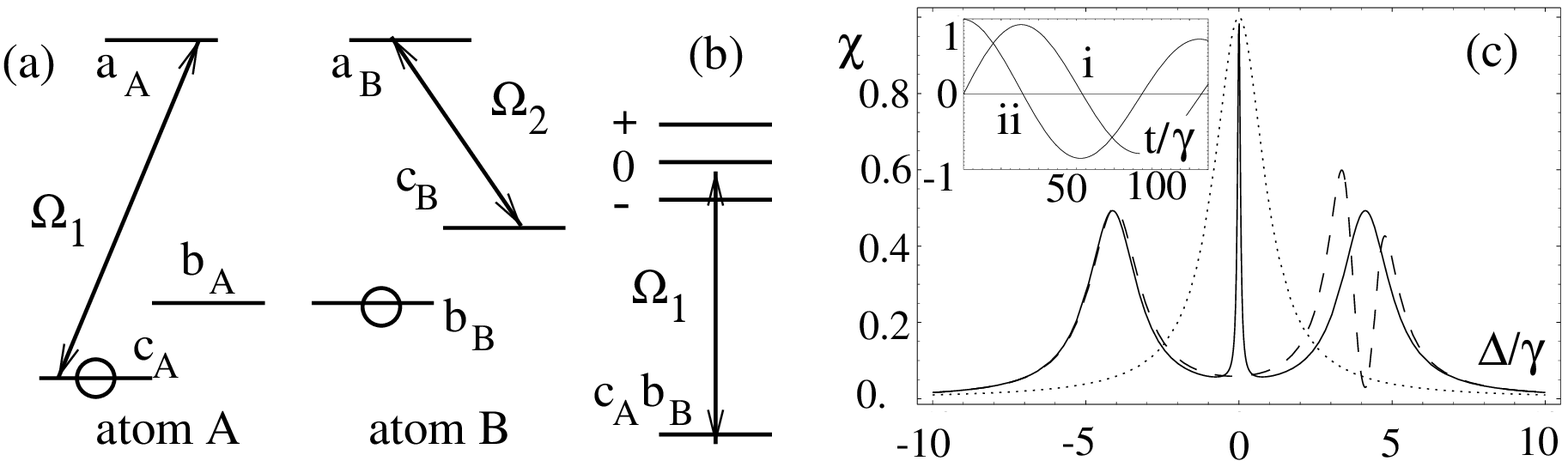,width=8.6cm}}
 \vspace*{2ex}
\caption{ (a) Schematic energy levels for two-atom Raman transitions. 
(b) Collective dressed states corresponding to Fig.1a.
(c) Susceptibility spectra (in arbitrary units) of a weak probe field
for the two-atom system of Fig.1a.  The dotted curve corresponds to absorption of
a free atom; for the solid curve $g = 4 \gamma,\nu_2 =\omega_{ac}^B, 
\omega_{ab}^A = \omega_{ab}^B,\Omega_2 = \gamma, \gamma_{bc}^k = 0$; for the dashed curve 
$\nu_2 -\omega_{ac}^B = 4 \gamma$. Inset shows the time 
evolution of the ground state coherences (i-$\sigma_{bc}^A$,
ii-$\sigma_{bc}^B/i$) 
corresponding 
to the system of Fig.1a excited at $t=0$ by a pair of identical fields
$\Omega_1 = \Omega_2 = \gamma$; $g = 20 \gamma$.} 

\end{figure}

%%%%%%%%%%%%%%%%%%%%%% figure 2 %%%%%%%%%%%%%%%%%%%%%%%%%%%%%%%%%%%%%%%%%%%%%

\begin{figure}[ht]
 \centerline{\epsfig{file=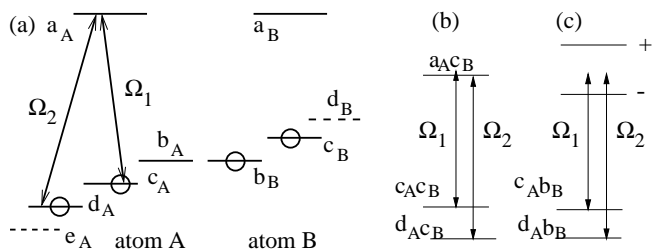,width=8.6cm}}
 \vspace*{2ex}
\caption{ (a) Schematic energy levels for conditional two-photon 
manipulations. (b) Corresponding two-atom states dressed by atom-atom 
interaction}

 \end{figure}

%%%%%%%%%%%%%%%%%%%%%% figure 3 %%%%%%%%%%%%%%%%%%%%%%%%%%%%%%%%%%%%%%%%%%%%%

\begin{figure}[ht]
 \centerline{\epsfig{file=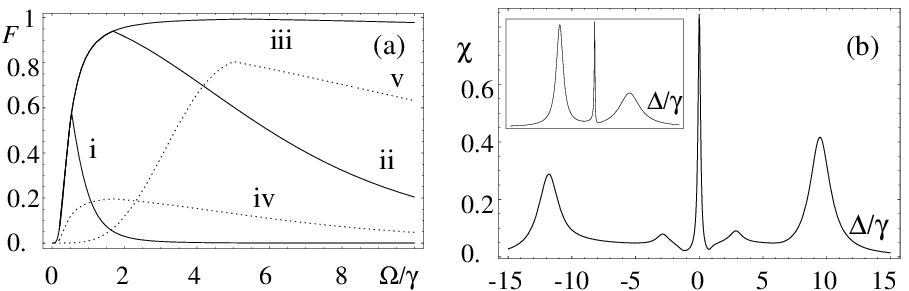,width=8.6cm}}
 \vspace*{2ex}
\caption{ (a) The minimum fidelity for conditional adiabatic passage as 
a function of the drive field Rabi frequency.  For curves (i-iii) 
$\gamma_{bc}^k = 0$, $T = 10 \gamma^{-1}$, and $g = 2 \gamma,20\gamma,
200\gamma$
respectively. For curves (iv,v)  $\gamma_{bc} = 0.1 \gamma$, $g = 20 \gamma$ 
and $T=10 \gamma^{-1}, 0.7 \gamma^{-1}$ respectively.
(b) Typical susceptibility spectra for cooperative Raman transitions in 
homogeneous mixture of atoms $A$ and $B$ with equal densities. 
$N \lambda^3 = 
10 \times (2\pi)^2 \gamma/\gamma_{rad}$. $\Omega = 3 \gamma,$ 
$\gamma_{bc}^k = 0.01 \gamma$. 
Inset demonstrates the influence of supper-radiant 
effects. All optical transitions are assumed to be radiatively 
broadened with identical decay rates. Other parameters correspond to 
those of the solid curve in Fig.1c. }

\end{figure}


\frenchspacing

\begin{references}

\bibitem{qcomp} See e.g. C.Williams and S. Clearwater, {\it Explorations
in Quantum Computing}, (Springer-Verlag, New-York, 1998); 
A.~M.~Stene, Rep. Prog. Phys. {\bf 61}, 117 (1998).
 

\bibitem{steve2} S.~E.~Harris, Physics Today, {\bf 50}(7), 36 
(1997).

\bibitem{hole} B.~Ham, M.~S.~Shahriar, and P.~Hemmer, Optics Lett.,
{\bf 22}, 1138 (1997); {\it ibid}, {\bf 24}, 86 (1999).

\bibitem{lene} L.~V.~Hau, S.~Harris, Z.~Dutton, and C.~Behroozi, 
Nature {\bf 397}, 594 (1999). 

\bibitem{fost} See e.g. Th. F{\"o}ster, in {\it Modern Quantum Chemistry},
O.~Sinanoglu Ed., (Academic Press, New-York, 1996). 

\bibitem{sau}  H. van Kampen {\it et al.}, Phys.Rev.A {\bf 56} 3569 (1997);
in a thermal vapor large collisional broadening will however tend to wash 
out the effects of interest here.

\bibitem{let} S.~K.~Sekatskii, and V.~S.~Letokhov, JETP Lett., 
{\bf 65}, 465 (1997). 


\bibitem{zoler} T.~Pelizzari, S.~A.~Gardiner, J.~I.~Cirac, and  P.~Zoller, 
Phys.Rev.Lett. {\bf 75}, 3788 (1995).

\bibitem{recent1} G.~Burennen, C.Caves, P. Jessen, and I. Deutsch, 
Phys.~Rev.~Lett. {\bf 82}, 1060 (1999); this implementation utilizes 
dipole-dipole interaction in a way different from the present 
approach.

\bibitem{recent2} D.~Jaksch, H.~Briegel, J.~I.~Cirac, C.~W.~Gardiner, and 
P.~Zoller, Phys.~Rev.~Lett. {\bf 82}, 1975 (1999). 


\bibitem{adiab} Alternatively, by tuning the optical 
fields such that dressed states $|0\rangle$ and $|\pm\rangle$ cross, 
a narrow transparency resonance (dashed curve in  Fig.1c) 
appears due to the generation of cooperative dark state of the type 
$|\Psi_{final}\rangle$. 
Adiabatic passage in this dark state can be 
also used for generation of entangled states and coherence transfer.


\bibitem{cstir} In order to perform quantum logic with conditional 
adiabatic passage 
an auxiliary state ($|e_A\rangle$ in Fig.2a) can be used. If this state is
initially empty, a conditional swapping of arbitrary amplitudes in 
the states $|c_A\rangle$)
and $|d_A\rangle$ can be achieved with a three-step transfer sequence 
$|c_A\rangle \rightarrow |e_A\rangle|$, $|d_A\rangle \rightarrow 
|c_A\rangle|$, $|e_A\rangle \rightarrow |d_A\rangle$. 

\bibitem{haroche} M.Gross, and S. Haroche, Phys.Reports {\bf 93}, 301 (1982).

\bibitem{bb} This is the result of an analysis based on an $N_A+N_B$ 
atom master equation, from which a hierarchy of evolution equations for
atomic correlations has been obtained. Spectrum in  Fig.3b is calculated
by making a Gaussian truncation of this hierarchy, and assuming a
homogeneous medium.  Detailed derivation and discussion of these results 
will be presented elsewhere.    

\bibitem{chuck} J.~Dowling and C.~Bowden,  Phys.~Rev.~Lett. {\bf 70}, 1421 (1993).


\bibitem{lat} E.g. the present technique can be used to 
alleviate stringent requirements on the precise location of atoms
in quantum logic gates based on optical lattices \cite{recent1}.


\bibitem{phil} Materials pertaining to spectral hole burning have recently 
been proposed for a cavity-QED implementation of quantum computing along the 
lines of Ref.\cite{zoler}; P.Hemmer, M.S.Shahriar, and S. Lloyd (unpublished).
 
\bibitem{YSO} R.W. Equall, R.L.Cone, R.M.Macfarlane 
Phys.Rev. B {\bf 47} 14741 (1993).

\bibitem{NV} 
X.-F. He, N.~Manson, and P.~Fisk, Phys.Rev.B {\bf 47} 8809 (1993); 
A.~Lenef {\it at al.}, {\it ibid}, {\bf 53} 13427 (1996).


\bibitem{singl} A.Gauber {\it et al.}, Science {\bf 276} 2012 (1997). 

\bibitem{cesium} S. I. Kanorsky, S. Lang, T. Eicher, K. Winkler, and A. 
Weis, Phys.Rev.Lett. {\bf81} 401 (1998).

\bibitem{t1} $T_1$s in some of these materials are on the order of minutes or 
longer. This suggests that it should be possible to significantly 
lengthen $T_2$ using techniques such as strong applied magnetic bias 
fields, milliKelvin temperatures, or NMR-based dephasing cancellation.
 

\end{references}
\end{document}